\documentstyle[11pt,aaspp4]{article}

\slugcomment{Submitted to the Astrophys. J. Letters}

\lefthead{Levshakov, Kegel \& Takahara}
\righthead{D/H at z = 3.57 toward Q 1937-1009}

\begin{document}

\title{ The D/H ratio at z = 3.57 toward Q 1937-1009 }

\author{Sergei A. Levshakov\altaffilmark{1}} 
\affil{National Astronomical Observatory, Mitaka, Tokyo 181, Japan}

\author{Wilhelm H. Kegel} 
\affil{Institut f\"ur Theoretische Physik der Universit\"at Frankfurt 
am Main, Postfach 11 19 32, 60054 Frankfurt/Main 11, Germany}

\and

\author{Fumio Takahara}
\affil{Department of Earth and Space Science, Faculty of Science,
Osaka University, Toyonaka, Osaka 560, Japan}

\altaffiltext{1}{On leave from
Ioffe Physico-Technical Institute, Russian Academy of Sciences} 

\begin{abstract}
Deuterium abundance re-measurements by 
Burles \& Tytler (1998; hereafter BT)
yielded D/H = (3.3$ \pm 0.3)\times10^{-5}$ and
the robust upper limit D/H $ < 3.9\times10^{-5}$ 
from the $z_{\rm a} = 3.572$ system toward Q~1937-1009. 
In this new analysis BT adopted multicomponent 
{\it microturbulent} models together with the
possibility to vary freely the local continuum 
level around each \ion{H}{1} line to improve the fit. 
The procedure failed, however, to fit adequately 
D~Ly-$\beta$ without recourse to an additional 
H~Ly-$\alpha$ contamination at the position of D~Ly-$\beta$.

We show that this obstacle may be successfully overcome within 
the framework of the {\it mesoturbulent} model accounting 
(in contrast to the microturbulent approximation) for a correlated 
structure of the large scale velocity field. 
Using the {\it same} observational data and 
the original continuum as determined by Tytler et al. (1996),
we obtained good fits.
The one-component mesoturbulent models provide D/H in the 
range $\simeq (3.2 - 4.8)\times10^{-5}$  
and the total hydrogen column density 
N(\ion{H}{1}) $\simeq (5.6 - 7.0)\times10^{17}$ cm$^{-2}$. 
This result is consistent with that found by us from
the $z_{\rm a} = 2.504$ and $z_{\rm a} = 0.701$ systems 
toward Q~1009+2956 and Q~1718+4807, respectively.
The range for D/H common to all three analyses is 
D/H $\simeq (4.1 - 4.6)\times10^{-5}$. This value is
consistent with standard big bang nucleosynthesis [SBBN] 
if the baryon-to-photon ratio, $\eta$, is in the range
$4.2\times10^{-10} \lesssim \eta \lesssim 4.6\times10^{-10}$, implying
$0.0155 \lesssim \Omega_{\rm b}h^2_{100} \lesssim 0.0167$.
\end{abstract}


\section{Introduction}

The most distant absorption-line system with observable 
\ion{D}{1} Ly-$\alpha$ and Ly-$\beta$ lines is  
the Lyman limit system discovered by Tytler et al. 
(1996, hereafter TFB) at $z_{\rm a} = 3.572$ toward the quasar 
Q~1937-1009 ($z_{\rm e} = 3.78$). In TFB, high resolution
(FWHM = 9 km s$^{-1}$) and high signal-to-noise 
(S/N $\simeq 75$ per pixel at the position of the Ly-$\alpha$ line)
Keck spectra of Q~1937-1009 revealed hydrogen absorption throughout
the entire Lyman series as well as a few metal absorption lines
with asymmetric profiles. Their first measurements of D/H based
on a two-component microturbulent model (Voigt profile
deconvolution analysis) gave a low value of 
D/H = $(2.3 \pm 0.3 \pm 0.3)\times10^{-5}$ ($1\sigma$  statistical and 
systematic errors). This result caused a lively discussion since
the low D/H value would imply a high universal 
density of baryons
$\Omega_{\rm b}h^2 = 0.024^{+0.006}_{-0.005}$ (TFB), 
where $\Omega_{\rm b}$ is the fraction of the critical density in
form of baryons and $h$ is the Hubble parameter scaled to 
100 km~s$^{-1}$~Mpc$^{-1}$. 
Note that the total SBBN baryon
density compatible with the abundances of D and $^3$He observed
in the solar system and
the interstellar medium, has been estimated to be 
$\Omega_{\rm b}h^2 = 0.0175^{+0.010}_{-0.005}$ (Hata et al. 1996),
which is slightly larger than earlier BBN estimates 
$\Omega_{\rm b}h^2 = 0.0125 \pm 0.0025$ (Walker et al. 1991).

The lack of uniqueness in the Voigt deconvolution procedure
was employed by Wampler (1996) to show that other
microturbulent  models could give D/H ratios that are about
3 times higher than the TFB's 
value and still are compatible with their fit.
Wampler also suggested that the total hydrogen column 
density may not be
well determined by TFB because of either incorrect sky subtraction
or improper modeling the Ly-$\alpha$ forest structure above and 
below the Lyman continuum break. The latter is problematical for
distant QSOs exhibiting a high density Ly-$\alpha$ forest, and
different methods of analysis have 
yielded by now different values for N(\ion{H}{1})~:
$(3.8 - 4.9)\times10^{17}$ cm$^{-2}$ by Songaila et al. (1997);
$(6.9 - 7.6)\times10^{17}$ cm$^{-2}$ by Burles \& Tytler (1997).
A value of N(\ion{H}{1}) $\sim 6\times10^{17}$ cm$^{-2}$ which
``produces a smooth forest opacity above and below the break''
was recently presented by Songaila (1998). 

Using new constraints on N(\ion{H}{1})$_{\rm tot}$, BT 
re-considered the
D/H measurements from the $z_{\rm a} = 3.572$ system. 
In this new approach, the metal absorption lines are not
used to constrain D/H.
The absorbers are separated into two groups -- with low and high
\ion{H}{1} column densities 
[N(\ion{H}{1}) $\leq 2.5\times10^{15}$ cm$^{-2}$ and
N(\ion{H}{1}) $\gtrsim 10^{17}$ cm$^{-2}$, respectively].
The first group does not show \ion{D}{1} and was utilized to fit
the Ly-$\alpha$ forest features in the vicinity of the 
Lyman series lines from the $z_{\rm a} = 3.572$ system.
The second (main) group of lines were described by five
free physical parameters~:    
N(\ion{H}{1}), N(\ion{D}{1}), $z$, $b_{\rm tur}$, and $T$.   
The best fitting model with 3 main components (Model~4)
allows for additional free parameters characterizing the 
local continuum to improve the fit. 
BT conclude, however, that  
``the model fit to Ly$\beta$ is not as good as Ly$\alpha$, 
there is under-absorption in two places near D-Ly$\beta$''.

The present Letter is primarily aimed at the inverse problem
in the analysis of the H+D Ly-$\alpha$ 
and Ly-$\beta$ absorption observed by TFB. 
It is shown that 
the mesoturbulent model based on only 5 physical parameters  
and an appropriate velocity field configuration
is a sufficient description of the
D~Ly-$\alpha$ and Ly-$\beta$ lines with
no free parameters in the continuum.

\section{Model assumptions and results}

It is generally believed  that the Lyman limit systems arise in
the outer regions of intervening galactic halos. 
We assume that the background ionizing radiation at
$z = 3.5$ is hard enough to keep the absorbing gas 
photoionized with $T_{\rm kin} \gtrsim 10^4$ K.
A galaxy halo is considered to be a continuous medium 
exhibiting a mixture of bulk motions such as infall and
outflows, rotation, tidal flows etc. 
The motion along the line of sight is then characterized 
by a fluctuating
velocity field which we consider as random. 
For the sake of simplicity, we assume a homogeneous 
(\ion{H}{1}-) density and temperature $T_{\rm kin}$. 
The model closely follows the papers by Levshakov \& Kegel
(1997; Paper~I), Levshakov, Kegel \& Mazets (1997; Paper~II)
and Levshakov, Kegel \& Takahara (1997; Paper~III); the 
reader is referred to these papers for more details. 
Here we only note that although the distribution function for the 
fluctuating velocity field
is assumed to be Gaussian {\it on average}, the resulting velocity
distribution along a {\it given} line of sight may deviate 
significantly from this.


To estimate physical parameters and an appropriate velocity field
structure along the sightline, we used a Reverse Monte Carlo [RMC]
technique described in Paper~III. The
algorithm requires to define 
a simulation box for the 5 physical parameters~:
N(\ion{H}{1}), D/H, $T_{\rm kin}$, 
$\sigma_{\rm t}/v_{\rm th}$, and $L/l$
[here $\sigma_{\rm t}$ denotes the {\it rms} velocity,
$v_{\rm th}$ the thermal width of the hydrogen lines,
$l$ the correlation length,
and $L$ the typical thickness of the absorbing region]. -- The 
continuous random function  $v(s)$ is represented by
its sampled values at equal space intervals $\Delta s$, i.e. by
the vector $\{v_1, v_2, \dots , v_k\}$ 
of the velocity components parallel to the line of sight
at the spatial points $s_j$ (see Paper~II).

In the present study we adopt for
the physical parameters the following boundaries~:
N(\ion{H}{1}) ranges from 
$3.8\times10^{17}$ to $7.5\times10^{17}$ cm$^{-2}$; 
D/H -- from
$2.5\times10^{-5}$ to $5.0\times10^{-5}$;
$T_{\rm kin}$ -- from $10^4$ to $2\times10^4$ K. 
For $\sigma_{\rm t}/v_{\rm th}$ the boundaries were set
from 1.0 to 4.0 to cover the
$\sigma_{\rm t}$-range estimated from 
the kinematic structure of the galactic halos observed at $z > 2$ 
[$\sigma_{\rm t} \simeq 40 \pm 15$ km s$^{-1}$,  
van Ojik  et al. (1997)]. Since for  
$L/l \gg 1$ the meso- and microturbulent profiles tend 
to be identical (Paper~I), 
we consider here only moderate $L/l$ ratios in the range 1.0~--~5.0.
Similar to BT, we assume that 
the D and H lines are not required to have 
the identical velocities as
the metal lines from the same system, and that
only the higher order Lyman series lines can trace the neutral
hydrogen distribution. 
But we fix $z_{\rm a}$ = 3.5723145 (the mean $z$ between the blue 
and the red components of the metal lines observed by TFB)
as a more or less arbitrary reference radial 
velocity at which $v_j = 0$.

Having specified the parameter space, we 
construct the objective function (similar to that in Paper~III) to
calculate $\chi^2_{\rm min}$. The objective function includes
the following portions of the Lyman series lines which
after preliminary analysis were chosen as most appropriate to
the simultaneous RMC fitting~: 
for H+D Ly-$\alpha$ and Ly-$\beta$,
$\Delta v $ ranges from $-133$ to $-39$ km~s$^{-1}$ 
(see Fig.~1{\bf a,b});
for Ly-14,\, 18 km~s$^{-1} \leq \Delta v \leq 42$ km~s$^{-1}$
(Fig.~1{\bf i});
for Ly-15, $\lambda\lambda = 4184.50 - 4185.00$ \AA\ 
(Fig.~1{\bf j}). 
In the Q~1937-1009 spectrum, there are many additional absorbers
blending the Lyman series lines. We do not fit these additional
absorptions since they do not affect significantly the D/H
measurement in this system as shown by BT. The strongest 
absorption feature seen 
at $z = 3.57295$ ($\Delta v = 41.7$ km~s$^{-1}$ in 
Fig.~1{\bf a-i}) has, according to BT, 
N(\ion{H}{1}) = $2.5\times10^{15}$ cm$^{-2}$,
which is less than 1\% of the total N(\ion{H}{1}) at $z = 3.57$.


The estimated parameters for a few adequate RMC profile fits 
are listed in Table~1.
We tried to find a satisfactory result with the reduced
$\chi^2$ per degree of freedom of 
$\chi^2_{\rm min} \lesssim \chi^2_{\nu,0.05}$ (where
$\chi^2_{\nu,0.05}$ is the expected $\chi^2$ value for 
$\nu$ degrees of freedom at the credible probability
of 95\%). In Table~1, the $\chi^2_{\rm min}$ values are shown
for the blue wings of the H+D Ly-$\alpha$ and Ly-$\beta$ lines
($\nu = 40$, $-133$~km~s$^{-1} \leq \Delta v \leq -39$ km~s$^{-1}$,
and $\chi^2_{40,0.05} = 1.40$), 
which are the only lines with the observable D-absorption.
Moreover, these pairs are more sensitive 
to the fitting procedure due to the highest S/N data obtained.
Slightly lower $\chi^2$ values were found in the combination with
the Ly-14 and Ly-15 lines which, however, have smaller weights
since their data are noisy. 
To illustrate our results, we show in Fig.~1 the best RMC
solution with the lowest $\chi^2_{\rm min}$ value (model {\bf d}).

To check the RMC solutions,
we calculated hydrogen profiles up to Ly-23 for each model and 
then superposed them to the corresponding parts of the 
Q~1937-1009 spectrum. For all models from Table~1 the results 
are similar to that shown in Fig.~1.
We do not find any pronounced discordance
of calculated and real spectra. 


The derived $v(s)$-configurations are not unique. 
Table~1 demonstrates the spread of the 
rms turbulent velocities from 
$\simeq 18$ km~s$^{-1}$  to $\simeq 22$ km s$^{-1}$. 
The projected velocity distribution functions $p(v)$ 
differ considerably 
from a Gaussian. Fig.~2 shows as an example for such
a distorted distribution 
(solid line histogram) the case of model {\bf d}.
By dotted lines we show three $p(v)$
for BT's best fit model with three components (Model~4).
Both the RMC $p(v)$ and the combined BT $p(v)$
distributions are asymmetric.
However, the RMC solution shows 
a stronger blue gradient as compared with BT.
This is the main reason why the absorption in 
the blue wing of the 
\ion{D}{1} Ly-$\beta$ line is enhanced without any
additional \ion{H}{1} interloper(s). 

\section{Conclusion}

We have shown that the interpretation of the Q~1937-1009 spectrum
obtained by TFB is not unique. The data can be modeled with a
higher D/H ratio if one accounts for spatial correlations
in the large scale velocity field.

The most accurate and 
robust RMC solution (model {\bf d}) 
was obtained with a total hydrogen
column density outside the range found by Burles \& Tytler (1997)
but in good agreement with the Songaila (1998) estimate.
Of course, our analysis gives a certain {\it range} for
N(\ion{H}{1})$_{\rm tot}$ and D/H (see Table~1). However,
this range narrows if we require the D/H value to be compatible
with the results of our previous analyses of the 
\ion{D}{1} absorption
toward Q~1009+2956 ($z_{\rm a} = 2.504$) (Paper~III) and
toward Q~1718+4807 ($z_{\rm a} = 0.701$) (Levshakov et al., 1998).
The range common to all three analyses is D/H 
$\simeq (4.1 - 4.6)\times10^{-5}$.
This implies for Q~1937-1009 N(\ion{H}{1}) 
$\sim (5.8 - 6.5)\times10^{17}$ cm$^{-2}$,
showing that model {\bf d} is well within this range.  
From SBBN it follows that D/H $\simeq (4.1 - 4.6)\times10^{-5}$
implies for the baryon-to-photon ratio, $\eta$, a value in
the interval $\simeq (4.2 - 4.6)\times10^{-10}$.
With the present-day photon density determined from the cosmic
microwave background (e.g. Fixen et al. 1996), one then estimates
the present-day baryon density to be in the range 
$\Omega_{\rm b}h^2 \simeq 0.0155 - 0.0167$.

The final conclusion is that the current observations
support SBBN and that there is no conflict with 
the D/H measurements within the generalized mesoturbulent
approach.

\acknowledgments

The authors are grateful to David Tytler 
for making available the calibrated Keck/HIRES echelle
spectra of Q~1937-1009 before their first publication
and acknowledge helpful correspondence
and comments by him, Xiao-Ming Fan and Scott Burles.
This work was supported in part by the RFBR grant No. 96-02-16905a.

\clearpage

\figcaption[]{ 
Observations (normalized flux) and RMC fits for Q~1937-1009.\,  
{\bf a-i} Velocity plots of 
the Keck/HIRES echelle data obtained by TFB
(dots and 1$\sigma$ error bars) and calculated RMC profiles  
convolved with the instrumental resolution of FWHM = 9~km~s$^{-1}$ 
(solid curves) corresponding to model {\bf d} in Table~1.
Zero velocity corresponds to the redshift $z = 3.5723145$ which
is the mean value between the blue and the red components of
the metal lines from Figure~1 in TFB. In panel {\bf a}, error
bars are too small to be distinguished. The under-absorption
on the blueward side of D~Ly-$\beta$ is within the statistically
expected scatter.\, {\bf j} Lyman-limit portion of the
Q~1937-1009 spectrum (dots and 1$\sigma$ error bars) and the
RMC solution (solid curve) corresponding to model {\bf d}.
\label{fig1} 
}

\figcaption[]{ 
Probability density function $p(v)$  of the best RMC solution shown
in Fig.~1 (solid line histogram) and 
for comparison three $p(v)$ distributions
with the velocity dispersions $b({\rm H})$ = 16.8, 18.9,
and 12.1 km~s$^{-1}$ (dotted curves)  
adopted from the best fit microturbulent Model~4 of BT.
The curves are weighted by the corresponding 
hydrogen column densities of 
$4.08\times10^{17}, 
2.15\times10^{17}$ and $1.38\times10^{17}$ cm$^{-2}$.
The sum of all three weighted functions is shown by the solid curve. 
\label{fig2} 
}

\clearpage

\begin{deluxetable}{ccccccc}
\footnotesize
\tablecaption{Cloud parameters derived from the Lyman series lines
by the RMC method. \label{tab1}} 
\tablewidth{16cm}
\tablehead{
\colhead{Model} & \colhead{N$_{17}$\tablenotemark{a}} & 
\colhead{D/H\tablenotemark{a}} & 
\colhead{$T_{4,\rm kin}$\tablenotemark{a}} & 
\colhead{$\sigma_t$\tablenotemark{a}} & 
\colhead{$L/l$} & \colhead{$\chi^2({\alpha,\beta})$}  
}
\startdata
({\bf a}) &5.60  & 4.78 & 1.16 &  20.1 & 3.6 & 1.24 \nl
({\bf b}) &5.94  & 4.41 & 1.43 &  22.3 & 3.5 & 1.29 \nl
({\bf c}) &6.04  & 4.10 & 1.09 &  20.8 & 2.8 & 1.40 \nl
({\bf d}) &6.15  & 4.23 & 1.15 &  20.0 & 3.6 & 1.18 \nl
({\bf e}) &6.85  & 3.74 & 1.38 &  18.5 & 2.0 & 1.39 \nl
({\bf f}) &7.00  & 3.20 & 1.54 &  18.3 & 2.1 & 1.45 \nl

\enddata
\tablenotetext{a}{
${\rm N}_{17}$ is the total hydrogen column density in
units of $10^{17}$ cm$^{-2}$, D/H in units of $10^{-5}$,
$T_{4,\rm kin}$ kinetic temperature
in units of $10^4$ K, 
$\sigma_t$ turbulent velocity in km~s$^{-1}$}

\end{deluxetable}

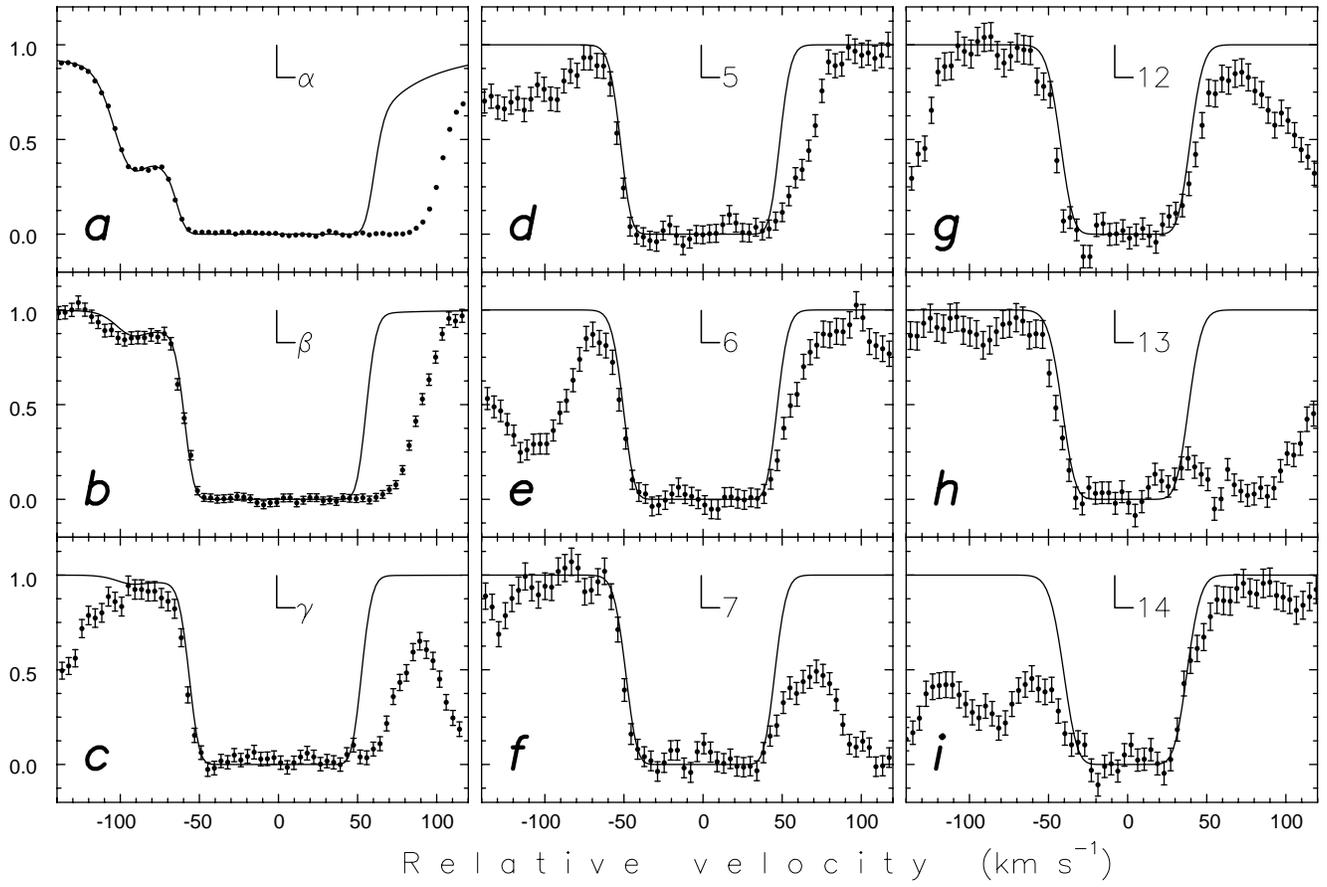
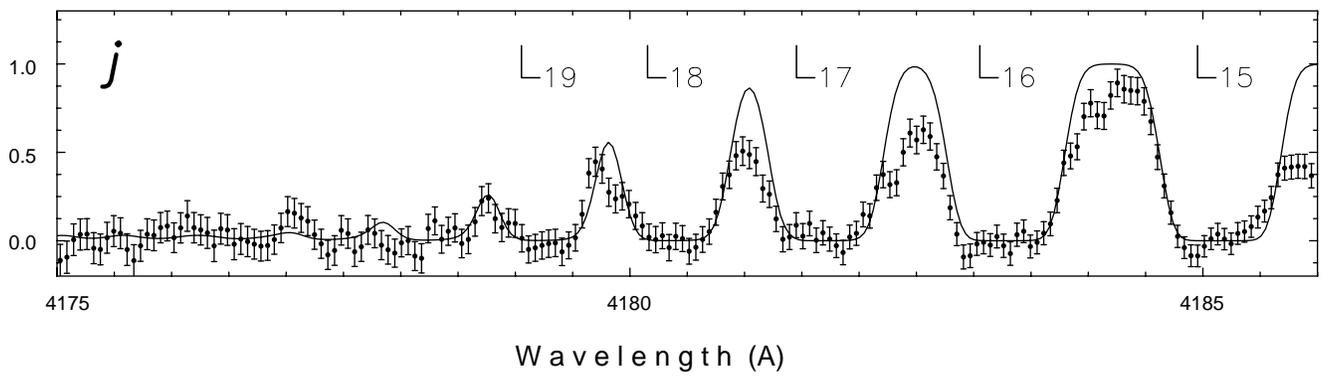
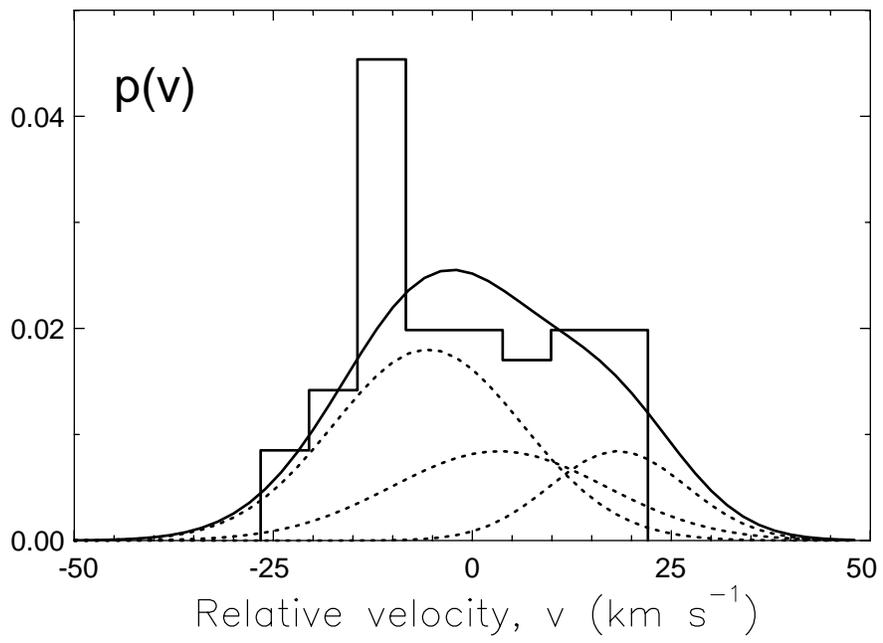
\end{document}